# TIME OF ARRIVAL BASED LOCALIZATION IN WIRELESS SENSOR NETWORKS: A LINEAR APPROACH

Ravindra S[1] and Jagadeesha S N[2]

[1,2] Jawaharlal Nehru National College of Engineering, Shimoga-577204.
Visvesvaraya Technological University,
Belguam, Karnataka,
India.
`rsp.cse@gmail.com`
`jagadeesha_2003@yahoo.co.in`

## ABSTRACT

*In this paper, we aim to determine the location information of a node deployed in Wireless Sensor Networks (WSN). We estimate the position of an unknown source node using localization based on linear approach on a single simulation platform. The Cramer Rao Lower Bound (CRLB) for position estimate is derived first and the four linear approaches namely Linear Least Squares (LLS), Subspace Approach (SA), Weighted Linear Least Squares (WLLS) and Two-step WLS have been derived and presented. Based on the simulation study the results have been compared. The simulation results show that the Two- step WLS approach is having higher localization accuracy.*

## KEYWORDS

*Source Localization, Time of Arrival, Linear Least Squares, Subspace Approach, Weighted Linear Least Squares (WLLS), Two-step WLS and CRLB*

## 1. INTRODUCTION

Wireless Sensor Network (WSN) consists of a large number of tiny low cost, low-power, multifunctional sensors which are capable of sensing, computing and communicating between these wireless devices which are deployed in a large geographic area[1]. WSN can be applied to a wide variety of diverse areas[2], such as environmental monitoring, military applications, target tracking, medical care, space exploration, location based services such as Emergency 911 (E-911) [3], Location sensitive billing, fraud detection, intelligent transport systems, location based Social Networking and Mobile yellow pages etc, [4]. Due to the developments in wireless communication WSN have been a new area of research [5-7]. Many applications of WSN require the sensor nodes to acquire the position information of the sensor nodes deployed. Data gathered by sensors should be associated with the sensors positions and it is worthless without the information about the place of its origin.

Despite the huge research effort, still a well accepted approach on how to solve the localization issue is being realized. Since the sensor nodes are inexpensive and are in huge number it is not practical to equip these sensors with a Global Positioning System (GPS) receiver. Various localization approaches have been proposed and can be seen in the literature [8-13] and there is not a single approach which is simple, distinct and gives decentralized solution for WSNs. The Ultra Wide Band (UWB) techniques [14] give very decent localization accuracy but the systems are expensive.

The commonly used approaches for measuring position estimate in WSN are Time of Arrival (TOA) [15], Time Difference of Arrival (TDOA)[16], Received Signal Strength (RSS)[17] and Angle of Arrival (AOA) a.k.a., Direction of Arrival (DOA)[18]. Where, the TOA, TDOA, and RSS measurement gives the distance calculation between the source sensor and the receiver sensors while DOAs provide the information of the angle and the distance measurements from the source and the receiver. Calculating these distance and angle measurements is not simple because of the nonlinear relationships with the source.

Given the TOA, TDOA, RSS and DOA information, the main focus of this paper is based on TOA positioning algorithms. We consider a two dimensional (2D) rectangular area where the sensors are deployed in Line-of-Sight (LOS) transmission, i.e., there is a direct path between the source and each receiver [19]. Also, we conclude that the measurements are well inside the expected range in order to obtain reliable location estimation.

The rest of the paper is organized as follows. In section 2 we present, the measurement model of TOA, and their positioning principles. In section 3, we provide the linear approach of finding the position by four methods i.e., Linear Least Squares (LLS), Subspace Approach (SA), Weighted Linear Least Squares (WLLS) and Two-step WLS. In section 4, the mean square position error comparison of the above approaches is made. Finally, the conclusions are drawn in section 5.

## 2. TOA Measurement Model and Principles of Source Localization

The mathematical measurement model for TOA based Source Localization Algorithm is given as:

$$\mathbf{r} = \mathbf{f}(x) + \mathbf{n} \quad (1)$$

Where $x$ is the source position which needs to be estimated, $\mathbf{r}$ the measurement vector, '$\mathbf{n}$' is an additive zero-mean noise vector and $\mathbf{f}(x)$ is a known nonlinear function in '$x$'.

### 2.1. Time of Arrival

TOA is the one-way propagation time of the signal travelling between a source and a receiver. This means that the source and all the receivers are accurately synchronized to measure the TOA information, and such an identical system is not needed if two way or round trip TOA is computed. The computed TOA is then multiplied with a known propagation speed, usually denoted as $c$, gives the measured distance between the source and the receivers. The measured TOA represents a circle with its centre at the receiver and the source must lie on the circumference in a Two Dimensional (2D) space.

Three or more such circles obtained from the noise free TOAs result in a distinct intersection point which represents the source position and is as shown in Figure 1(a) and 1(b), specifying that a minimum of three sensors is necessary for two dimensional position estimate [20]. If the number of sensors is less than three there is a possibility that there may not be any intersecting points and hence not a feasible solution. Hence, a minimum of three sensors is required to obtain the intersection and these can be represented as a set of circular equations, based on the optimization criterion the source position can be estimated with the knowledge of the known sensor array geometry [21, 22].

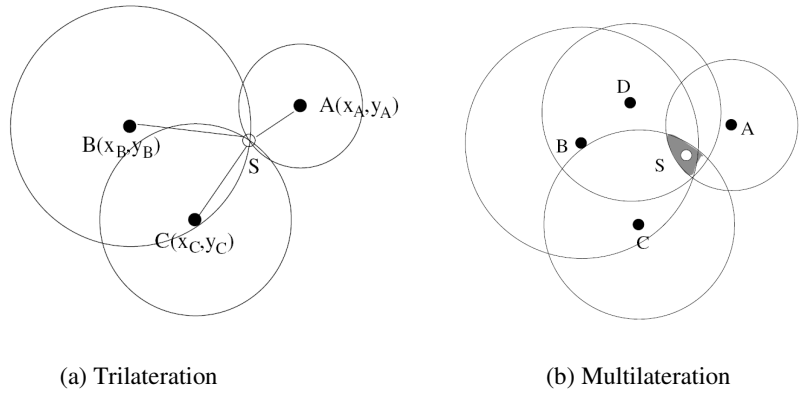

(a) Trilateration  (b) Multilateration

Figure 1: Geometrically representation of TOA Positioning system.

The TOA measurement model is developed as follows. Let $\mathbf{x}_l = [x_l \ y_l]^T$ be the known coordinates of the $l$ th sensor, $l = 1, 2, ......, L$. and $\mathbf{x} = [x \ y]^T$ be the unknown position of the source to be estimated. The number of receivers L must be greater than or equal to 3. The distance between the source and the sensor, denoted by $d_l$ is simply:

$$d_l = \|\mathbf{x} - \mathbf{x}_l\| = \sqrt{(x - x_l)^2 + (y - y_l)^2}, \quad l = 1, 2, .....L \tag{2}$$

The source radiates a signal at time 0 and the $l$ th sensor receives it at time $t_l$. That is, $\{t_l\}$ are the TOAs and is represented in a simple relationship between $t_l$ and $d_l$:

$$t_l = \frac{d_l}{c}, \quad l = 1, 2, ...., L \tag{3}$$

TOAs are prone to measurement errors. As a result, the range based measurement based on multiplying $t_l$ by c, denoted by $r_{\text{TOA},l}$ is modelled as:

$$r_{\text{TOA},l} = d_l + n_{\text{TOA},l} = \sqrt{(x - x_l)^2 + (y - y_l)^2} + n_{\text{TOA},l}, \quad l = 1, 2, .....L \tag{4}$$

where $n_{\text{TOA},l}$ is the range error in $r_{\text{TOA},l}$, which is resulted from TOA disturbance.

$$\mathbf{r}_{\text{TOA}} = \mathbf{f}_{\text{TOA}}(\mathbf{x}) + \mathbf{n}_{\text{TOA}} \tag{5}$$

where,

$$\mathbf{r}_{\text{TOA}} = [r_{\text{TOA}, 1} \ r_{\text{TOA}, 2} \ ........, \ r_{\text{TOA}, L}]^T \tag{6}$$

$$\mathbf{n}_{\text{TOA}} = [n_{\text{TOA}, 1} \ n_{\text{TOA}, 2} \ ........, \ n_{\text{TOA}, L}]^T \tag{7}$$

and

$$\mathbf{f}_{TOA}(\mathbf{x}) = \mathbf{d} = \begin{bmatrix} \sqrt{(x-x_1)^2 + (y-y_1)^2} \\ \sqrt{(x-x_2)^2 + (y-y_2)^2} \\ \vdots \\ \sqrt{(x-x_L)^2 + (y-y_L)^2} \end{bmatrix} \quad (8)$$

Here, $\mathbf{f}_{TOA}(\mathbf{x})$ represents the known function which is parameterized by $\mathbf{x}$ and in fact, it is the noise free distance vector. The source position estimation problem based on TOA measurements is to estimate $\mathbf{x}$ given $\{r_{TOA, l}\}$ or $r_{TOA}$.

A zero mean uncorrelated Gaussian process with variances $\{\sigma^2_{TOA, l}\}$ is assumed for the range error $\{n_{TOA,l}\}$. This helps in to facilitate the algorithm development and analysis as well as Cramer Rao Lower Bound (CRLB) Computation.

It is noteworthy that the zero – mean property indicates LOS transmission. The Probability Density Function (PDF) for each scalar random variable $r_{TOA, l}$, denoted by $p(r_{TOA, l})$, has the form of

$$p(r_{TOA,l}) = \frac{1}{\sqrt{2\pi\sigma^2_{TOA,l}}} \exp\left(-\frac{1}{2\pi\sigma^2_{TOA,l}}(r_{TOA,l} - d_l)^2\right) \quad (9)$$

And is characterized by its mean and variance, $d_l$ and $\{\sigma^2_{TOA, l}\}$, respectively. In other words, we can write

$$r_{TOA,l} \sim N(d_l, \sigma^2_{TOA,l}) \quad (10)$$

While the PDF for $r_{TOA}$, denoted by $p(r_{TOA})$, is

$$p(\mathbf{r}_{TOA}) = \frac{1}{(2\pi)^{L/2}|\mathbf{C}_{TOA}|^{1/2}} \exp\left(-\frac{1}{2}(\mathbf{r}_{TOA} - \mathbf{d})^T \mathbf{C}_{TOA}^{-1}(\mathbf{r}_{TOA} - \mathbf{d})\right) \quad (11)$$

where

$$\mathbf{C}_{TOA} = diag(\sigma^2_{TOA,1}, \sigma^2_{TOA,2}, \ldots, \sigma^2_{TOA,L}) \quad (12)$$

## 2.2. Cramer Rao Lower Bound

It is known that the MSPE of a biased estimator cannot be less than the CRLB. The mean square position error of the various positioning algorithms is computed and compared with CRLB which gives a lower bound on variance attainable by any unbiased estimators for the same data set [23]. Given the conditional PDF, we may derive the CRLB for TOA based location estimation and the same is given as follows [24, 25]:

- Calculate the second order derivatives of the logarithm of the measured PDF with respect to $\mathbf{x}$, that is $\partial^2 \ln p(\mathbf{r})/(\partial \mathbf{x} \partial \mathbf{x}^T)$..
- Take the expected value of $\partial^2 \ln p(\mathbf{r})/(\partial \mathbf{x} \partial \mathbf{x}^T)$.
- To yield $\mathbf{I}(\mathbf{x}) = E\{\partial^2 \ln p(\mathbf{r})/(\partial \mathbf{x} \partial \mathbf{x}^T)\}$

Where I(x) denotes the Fisher Information Matrix (FIM). And the lower bound for *x* and *y* are given by $\left[\mathbf{I}^{-1}(x)\right]_{1,1}$ and $\left[\mathbf{I}^{-1}(x)\right]_{2,2}$ respectively.

Alternatively, when the measurement errors are zero-mean Gaussian distributed, the FIM, whose elements are defined as:

$$\mathbf{I}(X) = \left[\frac{\partial \mathbf{f}(\mathbf{x})}{\partial \mathbf{x}}\right]^T \mathbf{C}^{-1} \left[\frac{\partial \mathbf{f}(\mathbf{x})}{\partial \mathbf{x}}\right] \tag{13}$$

where C is the covariance matrix. The FIM based on TOA measurements denoted by:

$$\mathbf{I}_{TOA}(X) = \left[\frac{\partial \mathbf{f}_{TOA}(\mathbf{x})}{\partial \mathbf{x}}\right]^T \mathbf{C}_{TOA}^{-1} \left[\frac{\partial \mathbf{f}_{TOA}(\mathbf{x})}{\partial \mathbf{x}}\right] \tag{14}$$

It is straightforward to show that

$$\left[\frac{\partial \mathbf{f}_{TOA}(\mathbf{x})}{\partial \mathbf{x}}\right] = \begin{bmatrix} \frac{x-x_1}{\sqrt{(x-x_1)^2+(y-y_1)^2}} & \frac{y-y_1}{\sqrt{(x-x_1)^2+(y-y_1)^2}} \\ \frac{x-x_2}{\sqrt{(x-x_2)^2+(y-y_2)^2}} & \frac{y-y_2}{\sqrt{(x-x_2)^2+(y-y_2)^2}} \\ \vdots & \vdots \\ \frac{x-x_L}{\sqrt{(x-x_L)^2+(y-y_L)^2}} & \frac{y-y_L}{\sqrt{(x-x_1)^2+(y-y_1)^2}} \end{bmatrix} \tag{15}$$

$$\left[\frac{\partial \mathbf{f}_{TOA}(\mathbf{x})}{\partial \mathbf{x}}\right] = \begin{bmatrix} \frac{x-x_1}{d_1} & \frac{y-y_1}{d_1} \\ \frac{x-x_2}{d_2} & \frac{y-y_2}{d_2} \\ \vdots & \vdots \\ \frac{x-x_L}{d_L} & \frac{y-y_L}{d_L} \end{bmatrix} \tag{16}$$

Employing eq. (16) and eq. (12), eq. (14) becomes

$$\mathbf{I}_{TOA}(\mathbf{x}) = \begin{bmatrix} \sum_{l=1}^{L} \frac{(x-x_l)^2}{\sigma_{TOA,l}^2 d_l^2} & \sum_{l=1}^{L} \frac{(x-x_l)(y-y_l)}{\sigma_{TOA,l}^2 d_l^2} \\ \sum_{l=1}^{L} \frac{(x-x_l)(y-y_l)}{\sigma_{TOA,l}^2 d_l^2} & \sum_{l=1}^{L} \frac{(y-y_l)^2}{\sigma_{TOA,l}^2 d_l^2} \end{bmatrix} \tag{17}$$

Where the lower bound for x and y are denoted by $\left[\mathbf{I}^{-1}(\mathbf{x})\right]_{1,1}$ and $\left[\mathbf{I}^{-1}(\mathbf{x})\right]_{2,2}$ respectively, and the

$CRLB_{TOA}(\mathbf{x})$, is

$$CRLB_{TOA}(\mathbf{x}) = \left[\mathbf{I}_{TOA}^{-1}(\mathbf{x})\right]_{1,1} + \left[\mathbf{I}_{TOA}^{-1}(\mathbf{x})\right]_{2,2} \tag{18}$$

## 3. LINEAR APPROACHES FOR SOURCE LOCALIZATION

Given the nonlinear expressions, the linear localization methodology tries to convert the nonlinear expressions of eq. (4), into a set of linear equations with zero mean disturbances. And a global solution is obtained based on the corresponding optimization cost function. In this paper four linear positioning approaches namely Linear Least Squares (LLS), Subspace Approaches (SA), WLLS and two Step WLS are presented.

### 3.1. Linear Least Squares

The LLS approach utilizes the ordinary Least Squares (LS) technique to estimate the position of $x$ by reorganizing eq. 4 into linear equations [26]. And an intermediate variable is added which is a linearization function to estimate the source position.

The linear TOA measurement model in **x** can be obtained by squaring eq. 4 on both sides,

$$r_{TOA,l}^2 = (x-x_l)^2 + (y-y_l)^2 + n_{TOA,l}^2 + 2n_{TOA,l}\sqrt{(x-x_l)^2 + (y-y_l)^2}, \tag{19}$$

where $l = 1,2,\ldots\ldots,L$
Let

$$m_{TOA,l} = n_{TOA,l}^2 + 2n_{TOA,l}\sqrt{(x-x_l)^2 + (y-y_l)^2} \tag{20}$$

be the noise term in eq. (19) and introduce a dummy variable $R$ of the form:

$$R = x^2 + y^2. \tag{21}$$

Substituting eq. (20) – eq. (21) into eq. (19) yields:

$$\begin{aligned} r_{TOA,l}^2 &= (x-x_l)^2 + (y-y_l)^2 + m_{TOA,l} \\ \Rightarrow r_{TOA,l}^2 &= x^2 - 2x_l x + x_l^2 + y^2 - 2y_l y + y_l^2 + m_{TOA,l} \\ \Rightarrow -2x_l x - 2y_l y + R + m_{TOA,l} &= r_{TOA,l}^2 - x_l^2 - y_l^2, \quad l = 1,2,\ldots,L \end{aligned} \tag{22}$$

Let

$$A = \begin{bmatrix} -2x_1 & -2y_1 & 1 \\ -2x_2 & -2y_2 & 1 \\ \vdots & \vdots & \vdots \\ -2x_L & -2y_L & 1 \end{bmatrix} \tag{23}$$

$$\boldsymbol{\theta} = [x \text{ y } R]^T \tag{24}$$

$$\boldsymbol{q} = \begin{bmatrix} m_{TOA,1} & m_{TOA,2} & \cdots & m_{TOA,L} \end{bmatrix}^T \tag{25}$$

and

$$\boldsymbol{b} = \begin{bmatrix} r_{TOA,1}^2 - x_1^2 - y_1^2 \\ r_{TOA,2}^2 - x_2^2 - y_2^2 \\ \vdots \\ r_{TOA,L}^2 - x_L^2 - y_L^2 \end{bmatrix} \tag{26}$$

The matrix form for eq. (22) is then:

$$\boldsymbol{A\theta} + \boldsymbol{q} = \boldsymbol{b} \tag{27}$$

where the observed $r_{TOA}$ of eq. (5) is now transformed to $\boldsymbol{b}$, $\boldsymbol{\theta}$ contains the source location to be determined and $\boldsymbol{A}$ is constructed from the known receiver positions. When $\{m_{TOA,l}\}$ are sufficiently small such that

$$\boldsymbol{q} \approx \begin{bmatrix} 2n_{TOA,1}\sqrt{(x-x_1)^2 + (y-y_1)^2} \\ 2n_{TOA,1}\sqrt{(x-x_2)^2 + (y-y_2)^2} \\ \vdots \\ 2n_{TOA,1}\sqrt{(x-x_L)^2 + (y-y_L)^2} \end{bmatrix} \tag{28}$$

can be considered a zero-mean vector, that is $E\{\boldsymbol{q}\} \approx 0$, we can approximate eq. (27) as:

$$\boldsymbol{A\theta} \approx \boldsymbol{b}. \tag{29}$$

The LS cost function based on eq. (29), denoted by $J_{LLS,TOA}(\tilde{\boldsymbol{\theta}})$, is:

$$\begin{aligned} J_{LLS,TOA}(\tilde{\boldsymbol{\theta}}) &= (\boldsymbol{A}\tilde{\boldsymbol{\theta}} - \boldsymbol{b})^T (\boldsymbol{A}\tilde{\boldsymbol{\theta}} - \boldsymbol{b}) \\ &= \tilde{\boldsymbol{\theta}}^T \boldsymbol{A}^T \boldsymbol{A} \tilde{\boldsymbol{\theta}} - 2\tilde{\boldsymbol{\theta}}^T \boldsymbol{A}^T \boldsymbol{b} + \boldsymbol{b}^T \boldsymbol{b} \end{aligned} \tag{30}$$

Which is a quadratic function in $\tilde{\boldsymbol{\theta}}$, and is a sole minimum in $J_{LLS,TOA}(\tilde{\boldsymbol{\theta}})$. The LLS estimate corresponds to:

$$\hat{\boldsymbol{\theta}} = \arg\min_{\tilde{\boldsymbol{\theta}}} J_{LLS,TOA}(\tilde{\boldsymbol{\theta}}) \tag{31}$$

which can be easily computed by differentiating eq. (30) with respect to $\tilde{\boldsymbol{\theta}}$ and equating the resulting expression to zero:

$$\left. \frac{J_{LLS,TOA}(\tilde{\boldsymbol{\theta}})}{\tilde{\boldsymbol{\theta}}} \right|_{\tilde{\boldsymbol{\theta}}=\hat{\boldsymbol{\theta}}} = 0$$

$$\Rightarrow 2\boldsymbol{A}^T\boldsymbol{A}\hat{\boldsymbol{\theta}} - 2\boldsymbol{A}^T\boldsymbol{b} = 0$$

$$\Rightarrow \boldsymbol{A}^T\boldsymbol{A}\hat{\boldsymbol{\theta}} = \boldsymbol{A}^T\boldsymbol{b}$$

$$\Rightarrow \hat{\boldsymbol{\theta}} = (\boldsymbol{A}^T\boldsymbol{A})^{-1}\boldsymbol{A}^T\boldsymbol{b} \tag{32}$$

The LLS position estimate can be obtained from the first and second entries of $\hat{\boldsymbol{\theta}}$, that is,

$$\hat{\boldsymbol{x}} = \left[ [\hat{\boldsymbol{\theta}}]_1 \; [\hat{\boldsymbol{\theta}}]_2 \right]^T \tag{33}$$

Eq. (33) is also known as the least squares calibration method [27].

### 3.2 Subspace Approach

The subspace positioning approach using TOA measurement is presented as follows. We first define a ($L$ x $2$) matrix **X:**

$$\boldsymbol{X} = \begin{bmatrix} x_1 - x & y_1 - y \\ x_2 - x & y_2 - y \\ \vdots & \vdots \\ x_L - x & y_L - y \end{bmatrix} \tag{34}$$

which is parameterized by x. With the use of X, the multidimensional similarity matrix [28], denoted by D, is constructed as:

$$\boldsymbol{D} = \boldsymbol{X}\boldsymbol{X}^T \tag{35}$$

whose $(m,n)$ entry can be shown to be

$$[\boldsymbol{D}]_{m,n} = (x_m - x)(x_n - x) + (y_m - y)(y_n - y)$$
$$= 0.5\left[(x_m - x)^2 + (y_m - y)^2 + (x_n - x)^2 + (y_n - y)^2 - (x_m - x_n)^2 - (y_m - y_n)^2\right]$$
$$= 0.5\left(d_m^2 + d_n^2 + d_{mn}^2\right) \tag{36}$$

where $d_{mn} = d_{nm} = \sqrt{(x_m - x_n)^2 + (y_m - y_n)^2}$ is of known value because it represents the distance between the $m$ th and $n$ th receivers. We then represent $\boldsymbol{D}$ using Eigen Value Decomposition (EVD):

$$\boldsymbol{D} = \boldsymbol{U}\Lambda\boldsymbol{U}^T \tag{37}$$

Where $U = [u_1\ u_2\ \cdots\ u_L]$ is an orthonormal matrix whose columns are corresponding eigen vectors and $\Lambda = diag(\lambda_1, \lambda_2, ......., \lambda_L)$ is the diagonal matrix of eigen values of D with $\lambda_1 \geq \lambda_2 \geq \cdots \geq \lambda_L \geq 0$. Noting that the rank of $D$ is 2, we have $\lambda_3 = \lambda_4 = \cdots \lambda_L = 0$. As a result eq. (37) can also be written as:

$$\begin{aligned} D &= U_S \Lambda_S U_S^T \\ &= U_S \Lambda_S^{1/2} (U_S \Lambda_S^{1/2})^T \\ &= U_S \Lambda_S^{1/2} \Omega (U_S \Lambda_S^{1/2} \Omega)^T \end{aligned} \quad (38)$$

Where $U_S = [u_1\ u_2]$, $\Lambda_S = diag(\lambda_1, \lambda_2)$ and $\Lambda_S^{\frac{1}{2}} = diag\left(\lambda_1^{\frac{1}{2}}, \lambda_2^{\frac{1}{2}}\right)$ denote the signal subspace components while $\Omega$ is the rotation matrix such that $\Omega \Omega^T = I$. Comparing eq. (35) and eq. (38) yields:

$$X = U_S \Lambda_S^{1/2} \Omega \quad (39)$$

We then determine the unknown rotation matrix

$$\begin{aligned} \Omega &= (U_S \Lambda_S^{1/2})^\dagger X \\ &= \left[(U_S \Lambda_S^{1/2})^T (U_S \Lambda_S^{1/2})\right]^{-1} (U_S \Lambda_S^{1/2})^T X \\ &= \Lambda_S^{-1/2} U_S^T X \end{aligned} \quad (40)$$

where $\dagger$ is moore penrose pseudo inverse.

Substituting eq. (40) into eq. (39) results in

$$X = U_S U_S^T X \quad (41)$$

which implies that the position **x** can be extracted from the eigen vectors of the signal subspace. As $d_l,\ l = 1, 2, ...., L,$ is not available, we construct a practical $D$ according to

$$[D]_{m,n} = 0.5(d_m^2 + d_n^2 + d_{mn}^2). \quad (42)$$

If the measurement error is present [28], $U_S \Lambda_S^{1/2}$ is the LS estimate of $X$ up to a rotation. And hence eq. (41) becomes an approximate relation.

To derive the position estimate, we first rewrite X as

$$X = Y - 1x^T \quad (43)$$

Where

$$\mathbf{Y} = \begin{bmatrix} x_1 & y_1 \\ x_2 & y_2 \\ \vdots & \vdots \\ x_L & y_L \end{bmatrix} \quad (44)$$

Using the subspace relation $U_S U_S^T = I - U_n U_n^T$ where $U_n = [u_3 \ u_4 \cdots u_L]$ corresponds to the noise subspace and Substituting eq. (43) into eq. (41), we get:

$$U_n U_n^T 1 x^T \approx U_n U_n^T Y \quad (45)$$

Following, the LLS procedure in eq. (29) – eq. (32), the subspace estimate of x using TOA measurements is computed as:

$$\hat{\mathbf{x}} = \left( \left( U_n U_n^T 1 \right)^\dagger U_n U_n^T Y \right)^T \quad (46)$$

$$= \frac{Y^T U_n U_n^T 1}{1^T U_n U_n^T 1} \quad (47)$$

The classical multidimensional scaling approach [29] is a modified subspace technique.

### 3.3 Weighted Linear Least Squares Approach

Since LLS is a simple approach and provides an optimum estimation performance only when the disturbances in the linear equations are independent and identically distributed. From eq. (28), it is clear that because of the noise vector q the LLS TOA-based positioning approach is suboptimal. The localization accuracy can be improved if we include a symmetric weighting matrix, say, **W**, in the cost function, denoted by $J_{WLLS,TOA}(\tilde{\theta})$. The final obtained expression is the WLS cost function which is of the form:

$$\begin{aligned} J_{WLLS,TOA}(\tilde{\theta}) &= (A\tilde{\theta} - b)^T W (A\tilde{\theta} - b) \\ &= \tilde{\theta}^T A^T W A \tilde{\theta} - 2\tilde{\theta}^T A^T W b + b^T W b \end{aligned} \quad (48)$$

According to eq. (27) – eq. (28), we have $E\{b\} = A\theta$ which corresponds to the linear unbiased data model. The optimum **W**, can be obtained similarly as best linear unbiased estimator (BLUE) [30, 31], which is equal to the inverse of the covariance of **q**. That is, the weighting matrix is similar to that if the maximum likelihood methodology. Employing eq. (28), we obtain:

$$\begin{aligned} W &= \left[ E\{qq^T\} \right]^{-1} \\ &\approx \left[ diag\left( 4\sigma_{TOA,1}^2 d_1^2, 4\sigma_{TOA,2}^2 d_2^2, \ldots, 4\sigma_{TOA,L}^2 d_L^2 \right) \right]^{-1} \\ &= \frac{1}{4} diag\left( \frac{1}{\sigma_{TOA,1}^2 d_1^2}, \frac{1}{\sigma_{TOA,2}^2 d_2^2}, \ldots, \frac{1}{\sigma_{TOA,L}^2 d_L^2} \right) \end{aligned} \quad (49)$$

As $\{d_l\}$ are not available, a small error condition can be obtained by replacing $d_l$ with $r_{TOA,l}$ which is valid an optimum $W$, for sufficiently small error condition:

$$W = \frac{1}{4} diag\left(\frac{1}{\sigma_{TOA,1}^2 r_{TOA,1}^2}, \frac{1}{\sigma_{TOA,2}^2 r_{TOA,2}^2}, \ldots \frac{1}{\sigma_{TOA,L}^2 r_{TOA,L}^2}\right) \quad (50)$$

Following eq. (31) – eq. (32), the WLLS estimate of $\theta$ is:

$$\hat{\theta} = \arg\min_{\tilde{\theta}} J_{WLLS,TOA}(\tilde{\theta})$$
$$= (A^T W A)^{-1} A^T W b \quad (51)$$

The Weighted Linear Least Square position estimate is given by eq. (33).

With only a moderate increase of computational complexity [31], eq. (51) is superior to eq. (32) in terms of estimation performance. The localization accuracy can be improved by making use of $[\hat{\theta}]_3$ according to the relation eq. (21) as follows. When $\hat{x}$ of eq. (51) is sufficiently close to $x$, we have:

$$[\hat{\theta}]_1^2 - x^2 = ([\hat{\theta}]_1 + x)([\hat{\theta}]_1 - x)$$
$$\approx 2x([\hat{\theta}]_1 - x) \quad (52)$$

Similarly, for $[\hat{\theta}]_2$:

$$[\hat{\theta}]_2^2 - y^2 \approx 2y([\hat{\theta}]_2 - y) \quad (53)$$

Based on eq. (21) and with the use of eq. (52) – eq. (53), we construct:

$$h = Gz + w \quad (54)$$

where

$$h = \left[[\hat{\theta}]_1^2 \; [\hat{\theta}]_2^2 \; [\hat{\theta}]_3\right]^T \quad (55)$$

$$G = \begin{bmatrix} 1 & 0 \\ 0 & 1 \\ 1 & 1 \end{bmatrix} \quad (56)$$

$$z = \left[x^2 \; y^2\right]^T \quad (57)$$

and

$$w = \left[2x([\hat{\theta}]_1 - x) \; 2y([\hat{\theta}]_2 - x) \; [\hat{\theta}]_3 - R\right]^T \quad (58)$$

23

Note that $z$ is the parameter vector to be determined. The result of BLUE is used to compute the covariance of $w$ in $\widehat{x}$ and is of the form [31]:

$$E\left\{\left[\left[\widehat{\theta}\right]_1 - x \ \left[\widehat{\theta}\right]_2 - y \ \left[\widehat{\theta}\right]_3 - R\right]\left[\left[\widehat{\theta}\right]_1 - x \ \left[\widehat{\theta}\right]_2 - y \ \left[\widehat{\theta}\right]_3 - R\right]^T\right\} = \left(A^T W A\right)^{-1} \quad (59)$$

Employing eq. (58) – eq. (59), the optimal weighting matrix for eq. (54), denoted by $\boldsymbol{\Phi}$, is then:

$$\boldsymbol{\Phi} = \left[diag(2x, 2y, 1)\left(A^T W A\right)^{-1} diag(2x, 2y, 1)\right]^{-1} \quad (60)$$

As a result, the WLLS estimate of $z$ is

$$\widehat{z} = \left(G^T \boldsymbol{\Phi} G\right)^{-1} G^T \boldsymbol{\Phi} h \quad (61)$$

As there is no sign information for $x$ in $z$, the final position estimate is determined as:

$$\widehat{x} = \left[\text{sgn}\left(\left[\widehat{\theta}\right]_1\right)\sqrt{[\widehat{z}]_1} \ \ \text{sgn}\left(\left[\widehat{\theta}\right]_2\right)\sqrt{[\widehat{z}]_2}\right]^T \quad (62)$$

Where sgn represents the signum function [32]. This technique is called the two-step WLS estimator [33] where eq. (21) is used in an implicit manner. Similarly an explicit way is to use Lagrangian multipliers [34, 35] to minimize eq. (48) subject to the constraint of eq. (21).

## 4. SIMULATION RESULTS

The performance evaluation of the various linear TOA based localization approaches is simulated using MATLAB[TM] Version 7.10.0.499 (R2010A) on Microsoft Windows XP®, Professional Version 2002, Service Pack 3, 32 bit operating system installed on Intel[R], Core[TM] 2 Duo CPU, E4500 @ 2.20GHz, 2.19GHZ, 2.0Gb of Ram.

The simulation is done in a 2- Dimensional region with a size of (1100m x 1100m), where the unknown source is assumed to be at position (x, y) = (200, 300), and the receivers are positioned in known coordinates at (0, 0), (0, 1000), (1000, 1000) and (1000, 0) respectively. And the sensors are deployed in a rectangular area where the source is surrounded by four receivers and is shown in Figure 2. It is also assumed that $\{n_{TOA,l}\}$ are zero- mean uncorrelated Gaussian process with variances $\{\sigma_{TOA,l}^2\}$, and zero- mean property indicates LOS transmission. The range error variance $\sigma_{TOA,l}^2$ is proportional to $d_l^2$ with $SNR = d_l^2 / \sigma_{TOA,l}^2$. The signal-to- noise-ratio (SNR) = 30dB has been assumed. All the methods estimate the position. The implementation flow of the two step WLS and the CRLB is shown in Figure 3 and 4 respectively. Figure 5, shows the plot of Mean Square Position Error (MSPE) defined as $E\left\{(\widehat{x}-x)^2 + (\widehat{y}-y)^2\right\}$ of different linear approaches and CRLB for SNR in the range [-10, 25] dB, based on 1000 independent runs, which is given by $\sum_{i=1}^{1000}\left[(\widehat{x}_i - x)^2 + (\widehat{y}_i - y)^2\right]/1000$, where $(\widehat{x}_i, \widehat{y}_i)$, denotes the position estimate of the $i$ th run. From the Figure, it can be seen that the Two-step WLS estimator achieves the optimal estimation performance, while the LLS, SA and WLLS approaches are suboptimal.

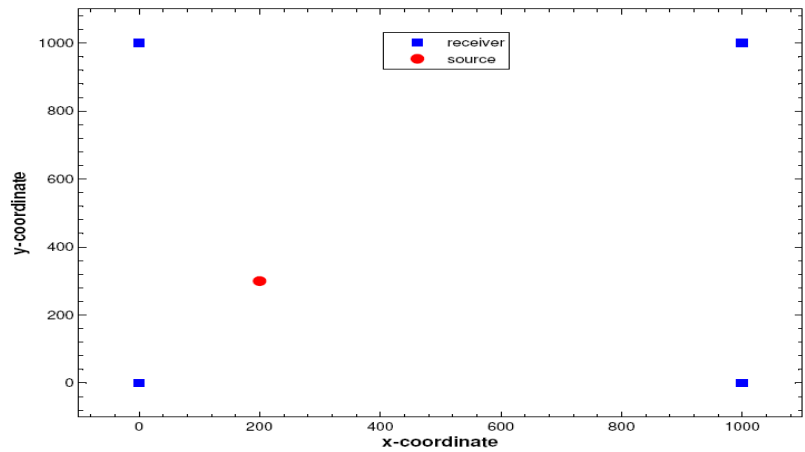

Figure 2: Position of Source and Receivers

Table 1: Estimated TOA positions obtained using different linear approaches.

| Method | $\hat{x}$ in meters | $\hat{y}$ in meters |
|---|---|---|
| **Linear Least Squares** | 201 | 302 |
| **Subspace Approach** | 197 | 297 |
| **Weighted Linear Least Squares** | 199 | 300 |
| **Two Step Weighted Least Squares** | 200 | 300 |

Table 1 gives the results of the position estimate, and results of the two step WLS approach is better when compared with the other linear approaches. The two step WLS accurately estimates the positions. The accuracy is higher in case of Two- step WLS while the other approaches have lower localization accuracy.

25

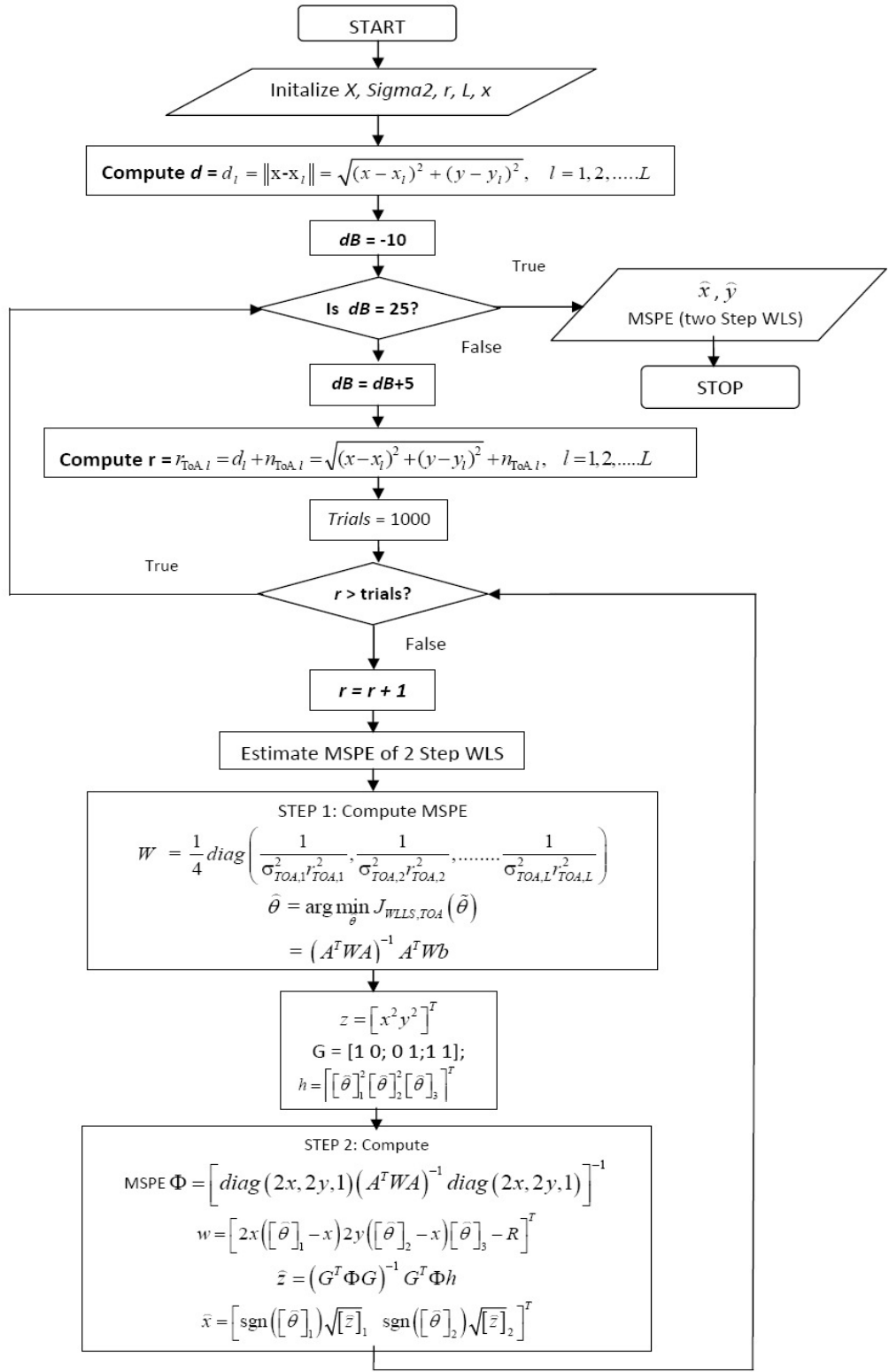

Figure 3: Flowchart showing the computation of Two- Step WLS approach

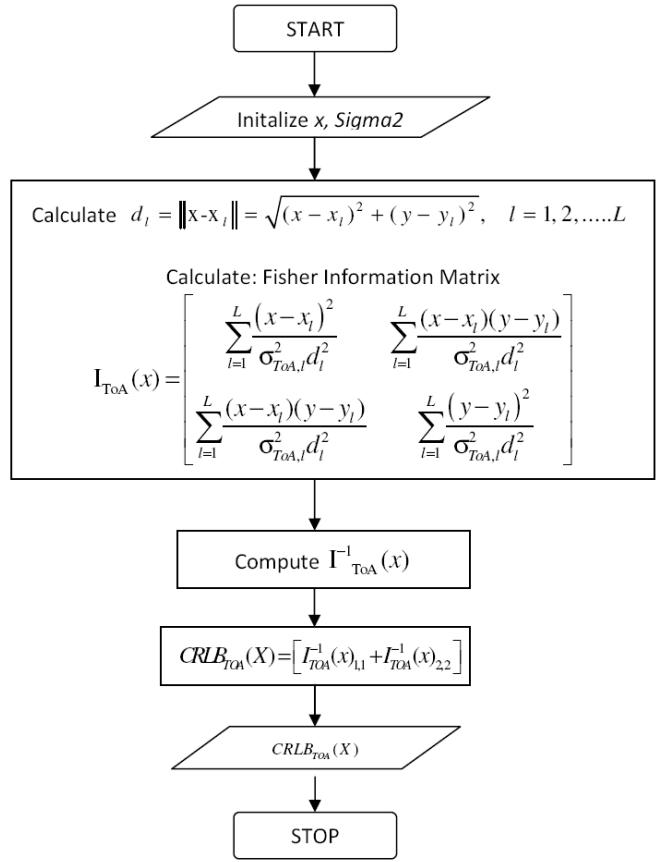

Figure 4: Flowchart showing the computation of CRLB

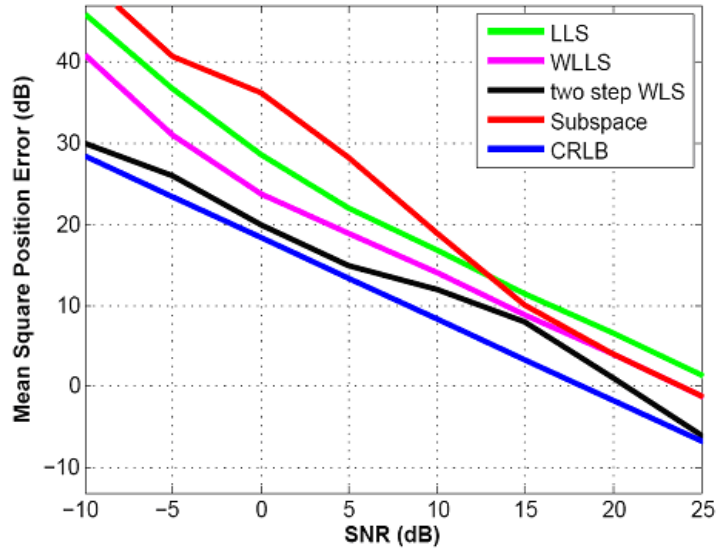

Figure 5: Mean Square Position Error Computation of the different linear based approaches

27

## 5. CONCLUSIONS

The work presented addresses the problem of position estimation of a sensor node in a Wireless Sensor Network, using TOA measurements in LOS environments. The CRLB for the position estimation problem has been derived first and later four methods namely LLS, SA, WLLS and Two Step WLS methods of linear approach have been derived and presented. Extensive simulations have been carried out and the results of different methods have been compared. The comparison reveals that the two step WLS method is superior to the rest of the linear approaches in LOS environments.

We have restricted our studies to the linear approaches. The work can be extended to the nonlinear approaches also and shall be reported in a future communication. Tracking mobile nodes is an interesting problem which may require a combination of two or more approaches to improve the accuracy of the position estimate.

## ACKNOWLEDGEMENTS

The authors would like to thank Dr. Yerriswamy T, (PDIT Hospet), and their Colleagues at Dept. of CSE, JNNCE Shimoga, for the valuable inputs and the reviewers for their useful comments and suggestions.

**Authors**

Ravindra. S. received his B.E., in Electrical and Electronics Engineering., and M.Tech., in Networking and Internet Engineering, from Visvesvaraya Technological University, Belgaum, Karnataka, India in 2006 and 2008 respectively. He is currently working towards a Doctoral Degree from Visvesvaraya Technological University, Belgaum, Karnataka, India. At present he is working as Assistant Professor, in Computer Science and Engineering department of Jawaharlal Nehru National College of Engineering (affiliated to Visvesvaraya Technological University), Shimoga, Karnataka, India.

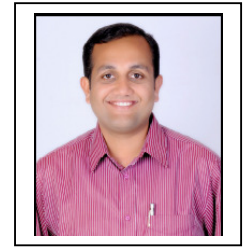

Dr. S.N. Jagadeesha received his B.E., in Electronics and Communication Engineering, from University B. D. T College of Engineering., Davangere affiliated to Mysore University, Karnataka, India in 1979, M.E. from Indian Institute of Science (IISC), Bangalore, India specializing in Electrical Communication Engineering., in 1987 and Ph.D. in Electronics and Computer Engineering., from University of Roorkee, Roorkee, India in 1996. He is an IEEE member. His research interest includes Array Signal Processing, Wireless Sensor Networks and Mobile Communications. He has published and presented many papers on Adaptive Array Signal Processing and Direction-of-Arrival estimation. Currently he is professor in the department of Computer Science and Engineering, Jawaharlal Nehru National College of Engg. (Affiliated to Visvesvaraya Technological University), Shimoga, Karnataka, India.

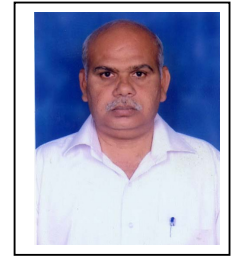